\documentclass{article}

\usepackage{arxiv}

\usepackage{hyperref}       
\usepackage{xcolor}
\hypersetup{colorlinks=true, linkcolor=blue, citecolor=blue}
\usepackage{amsfonts}       
\usepackage{amsmath}
\usepackage{amssymb}
\usepackage{nicefrac}       
\usepackage{microtype}      
\usepackage{lipsum}         
\usepackage{graphicx}
\usepackage{natbib}
\usepackage{doi}
\usepackage{siunitx}
\newcommand{\bm}[1]{{\mbox{\boldmath $#1$}}}

\title{Flow control by a hybrid use of machine learning and control theory\thanks{
This is an original manuscript of an article published by Emerald Publishing Limited in {\it Int.~J.~Numer.~Meth.~Heat~Fluid~Flow} on 12 Aug 2024, available online:
https://doi.org/10.1108/HFF-10-2023-0659}}

\newif\ifuniqueAffiliation
\uniqueAffiliationtrue

\ifuniqueAffiliation 

\author{Takeru Ishize \\
	Department of Mechanical Engineering,
	Keio University\\
	takeru.ishize@keio.jp \\
	\And
  Hiroshi Omichi\\
  Department of Mechanical Engineering,
	 Keio University;\\
	 Department of Mechanical and Aerospace Engineering,
	 UCLA\\
	 hiroshi.omichi@keio.jp\\
	\And
 Koji Fukagata \\
	Department of Mechanical Engineering,
	Keio University\\
	fukagata@mech.keio.ac.jp\\
}

\else
\usepackage{authblk}

\newbox{\orcid}\sbox{\orcid}{\includegraphics[scale=0.06]{orcid.pdf}} 
\author[1]{%
	Takeru Ishize 
	\texttt{takeru.ishize@keio.jp} }
\author[1,2]{%
	\href{https://orcid.org/0000-0000-0000-0000}{\usebox{\orcid}\hspace{1mm}Elias D.~Striatum\thanks{\texttt{stariate@ee.mount-sheikh.edu}}}%
}
\affil[1]{Department of Mechanical Engineering, Keio University, Japan}\\
\fi

\renewcommand{\headeright}{}
\renewcommand{\undertitle}{}
\renewcommand{\shorttitle}{
This is an original manuscript of an article published by Emerald Publishing Limited in {\it Int.~J.~Numer.~Meth.~Heat~Fluid~Flow} on 12 Aug 2024, available online:
https://doi.org/10.1108/HFF-10-2023-0659}

\begin{document}
\maketitle


\begin{abstract}

\noindent {\bf Purpose}
Flow control has a great potential to contribute to the sustainable society through mitigation of environmental burden.
However, high dimensional and nonlinear nature of fluid flows poses challenges in designing {efficient control laws using the control theory.}
This paper aims to propose a hybrid method ({\it i.e.}, machine learning and control theory) for feedback control of fluid flows{, by which the flow is mapped to the latent space in such a way that the linear control theory can be applied therein.}

\noindent {\bf Design/methodology/approach}
{We propose a partially nonlinear linear system extraction 
autoencoder (pn-LEAE), which consists of convolutional neural networks-based autoencoder (CNN-AE) and a custom layer to extract a low-dimensional latent dynamics
from fluid velocity field data.
This pn-LEAE is designed to extract a linear dynamical system so that the modern control theory can easily be applied, while a nonlinear compression is done with the autoencoder (AE) part so that the latent dynamics conforms to that linear system.
The key technique is to train this pn-LEAE with the ground truths at two consecutive time instants, whereby the AE part retains its capability as the AE, and the weights in the linear dynamical system are trained simultaneously.}

\noindent {\bf Findings}
We demonstrate the effectiveness of the linear system extracted by the pn-LEAE, as well as the designed control law's effectiveness for a flow around a circular cylinder at the Reynolds number of ${\rm Re}_{D}=100$.
{When the control law derived in the latent space was applied to the direct numerical simulation, the lift fluctuations were suppressed 
{over}
50\%.}

\noindent {\bf Originality/value}
This is the first attempt utilizing CNN-AE for linearization of fluid flows involving transient development to design a feedback control law.


\noindent {\bf Paper type} Research paper\\

\end{abstract}

\baselineskip 16pt

\section{Introduction}
Flow control has always been a significant focus in fluid mechanics,
aiming at improving flow characteristics such as reducing drag, enhancing mixing, and minimizing noise, and 
the ability to manipulate fluid flow in a versatile manner is not only a matter of scientific interest but also holds substantial engineering importance.
Flow control strategies can be broadly classified into passive flow control (PFC) and active flow control (AFC). 
PFC typically involves geometric modifications within the initial setup that do not require external energy input in its operation. 
In contrast, AFC employs actuators and requires external energy input in its operation.
AFC is gaining increased attention due to its potential for delivering more pronounced effects over a broader operational range than PFC.  
{AFC can further be classified into predetermined ({\it i.e.}, open-loop) control and  feedback ({\it i.e.}, closed-loop) control.
Predetermined control laws have often been proposed based on the physical insight on the flow phenomena, such as
uniform blowing \citep{Kornilov2015}, spanwise wall motion \citep{Karniadakis:03,Ricco:21}, and streamwise traveling waves of wall-normal forcing \citep{fukagata2024}, to name a few.
Feedback control laws have also often been proposed based on the physical insight \citep{Kim2003}; but at the same time,
applications of the modern control theory have also extensively been investigated since 1990s 
due to its great potential
\citep{abergel1990,roussopoulos1993feedback,kim2007linear}.}
{To derive a feedback control law using the modern control theory, we usually consider
a state equation such as
\begin{align}
\frac{d\bm{r}}{dt}&=\bm{N}(\bm{r},\bm{m}),
\label{eq:system}
\end{align}
where the nonlinear function $\bm{N}$ representing the system dynamics depends on the state $\bm{r}$ and the control input $\bm{m}$; then, the control input $\bm{m}$ is obtained so as to minimize or maximize a prescribed cost function $J$.}

There are primarily two categories of control theories in AFC: model-based methods and model-free methods. 
Model-free methods derive control laws from data in an unsupervised manner, enabling their application without the need for modeling phenomena \citep{brunton2015closed}. 
These methods find extensive use across numerical and experimental systems. 
In recent times, machine learning (ML) techniques, including reinforcement learning and genetic programming, have been used to discover control laws \citep{brunton2020machine}, demonstrating significant potential for AFC \citep{rabault2019artificial,ren2019active}. 
{Nevertheless, ML-based control methods are confronted with the challenge of the black-box problem. 
While several studies have endeavored to tackle this issue~\citep{Castellanos2022,Hou2023}, 
it remains as a major area for
 further investigation.
Moreover, numerous unsolved challenges persist in model-free control approaches, including limitations in control efficacy, interpretability of control mechanisms, and reliability.
}

In contrast, model-based methods derive control laws from system dynamics {({\it i.e.}, Eq.~(\ref{eq:system}))} using mathematical theories, encompassing linear and stochastic models, among others.
{In phenomena amenable to modeling, model-based methods are generally acknowledged for their ease of handling, performance, analytical tractability, and high reliability.}
These methods have been applied to numerous flow control problems \citep{Rowley2006,kim2007linear,Pastoor2008,choi2008,kasagi2009microelectromechanical}. 
However, modeling flow phenomena remains difficult due to their high dimensionality and strong nonlinearity, and their complexity has prevented the establishment of universal control methods.
{
In fact, we encounter this difficulty even for trivial flow problems such as a two-dimensional periodic vortex shedding around a circular cylinder at low Reynolds numbers.
For instance, \cite{illingworth}, who modeled the flows around a circular cylinder using the eigensystem realization algorithm and attempted to suppress vortex shedding, demonstrated that the linear feedback control does not work
already at the Reynolds number ${\rm Re}_D>80$ due to the disappearance of the gain window to stabilize the unsteady modes.
}
Therefore, there is a strong need to develop a novel framework to address these intricate issues.
{We anticipate that the enhancement in modeling capabilities to represent fluid phenomena will directly translate into a substantial improvement in the efficacy of model-based control strategies.}

To address these challenges, the utilization of ML techniques is under extensive investigation.
ML methods are data-driven and are renowned for their adaptability across diverse cases, without necessitating mathematical models of the target systems.
Machine learning techniques have found broad applications in the field of fluid mechanics~\citep{brunton2020machine,FFT2020,SLB2021}, including state estimation~\citep{Milano2002,tong2022estimating,farazmand2023tensor}, reduced-order modeling (ROM)~\citep{Guastoni2021,THBSDBDY2020}, enhancement in numerical simulations and experiments~\citep{simu2021,Vinuesa2022,Vinuesa2023}, turbulence modeling~\citep{DIX2019}, and super-resolution reconstruction~\citep{fukami2023,Guemes2022} to name a few.
Among these, reduced-order modeling has gained significant attention because of its capability to efficiently handle large datasets.
In particular, the autoencoder (AE) \citep{hinton2006reducing}, which maps high dimensional fluid data into a low-dimensional feature space, is widely utilized.
AE is an unsupervised model that can be trained using the same data before and after the bottleneck structure to extract low-dimensional features.
\citet{hasegawa2020cnn} applied an AE-based ROM to flows around cylinders at various Reynolds numbers, demonstrating its applicability.
\citet{san2018extreme} introduced an extreme learning-based ROM for quasi-stationary geophysical turbulent flows and demonstrated its superiority in terms of stability when compared to the proper orthogonal decomposition (POD)~\citep{lumley1967structure}-based ROM.
Various other studies have reported the use of AE-based ROMs \citep{brenner2019perspective,brunton2020special,fukagata2023}.
Consequently, the utilization of the AE-based approach for high-dimensional fluid data has demonstrated the wide applicability of latent variables within the extracted low-dimensional space.

However, the challenge of strong nonlinearity remains unresolved, and the low-dimensional latent vectors obtained also exhibit nonlinearity.
For instance, \citet{fukami2021sindy} low-dimensionalized a flow around a circular cylinder at the Reynolds number 
${\rm Re}_D=100$ using a Convolutional Neural Network (CNN)~\citep{LBBH1998}-based AE (CNN-AE) and extracted the ordinary differential equation (ODE) governing the low-dimensional latent dynamics using SINDy \citep{sindy}.
While the dimension of the latent space was reduced to two, the obtained ODE eligible of capturing the transient flow was a set of fifth-order nonlinear equations, which is difficult to apply the conventional control theory.
{Moreover, \citet{wang2024} have recently proposed a methodology for modeling flow dynamics in the latent space using the $\beta$-variational autoencoder ($\beta$-VAE) \citep{kingma2013,higgins2017}. 
Nevertheless, the utilization of $\beta$-VAE essentially reduces complex high-dimensional flow data to stochastic latent variables without adherence to explicit governing equations, thus raising concerns regarding the applicability of 
control 
theory.
}
Therefore, the present study aims to develop a machine learning-based method, referred to as
a linear system extracti{on}
autoencoder (LEAE), to extract 
a
linear dynamical system
in the latent space.
Although the structure of the LEAE is mostly similar to the conventional CNN-AE, the LEAE is designed to output the flow field at the next time step ($t+\Delta t$) from that at the current time step ($t$).
For that purpose, the LEAE incorporates a custom layer 
which performs a time integration of the latent dynamics
based on the Crank-Nicholson scheme. 

While the LEAE is expected to simplify the problem by
mapping the high-dimensional flow field governed by a nonlinear dynamics to a
low-dimensional latent 
variables governed by a linear dynamics,
always assuming linear systems 
poses difficulties
when
the nonlinearity plays an essential role, such as in the cases of transient flows.
To handle this issue,
we also propose an enhanced version of LEAE, called a Partially Nonlinear Linear System Extraction Autoencoder (pn-LEAE), which can account for
a nonlinear variation of the orbital radius of the latent variables.
This pn-LEAE, focusing on continuously changing flow fields, extracts a system capable of representing the time evolution of latent variables.
In the present study, as a proof of concept, 
we 
consider
two types of flows: 1) transient flow and 2) steady periodic flow around a circular cylinder at $Re_D=100$. 
This kind of low-Reynolds number flow around a circular cylinder has long been considered in the studies of feedback control aiming at suppression of vortex shedding
\citep{park1994feedback};
however, {as mentioned above}, 
the
linear feedback control does not work
{already} at ${\rm Re}_D>80$ 
due to disappearance of the gain window \citep{illingworth}.

The paper is organized as follows.
In Section 2, the method for preparing the dataset, the details of the proposed LEAE, and the method of flow control using the LEAE are introduced.
In Section 3, we assess the linear system extracted through the 
proposed
procedure and demonstrate its effectiveness. 
Then, we design a control law based on the extracted linear system and investigate its performance.
Finally, conclusions are drawn in Section 4.

\section{Methods}
\subsection{Dataset}
\label{subsec:DNS}
We 
consider
the two-dimensional (2D) flow around a circular cylinder as a basic example.
Training data for machine learning (ML) model is prepared by direct numerical simulation (DNS)~\citep{DNS_2D}, whose
governing equations are the incompressible continuity and Navier-Stokes equations, {\it i.e.},
\begin{align}
    {\bm \nabla} \cdot \bm u &= 0,\\
\frac{\partial \bm{u}}{\partial t}+{\bm \nabla} \cdot (\bm{uu})&=- {\bm \nabla} p+\frac{1}{{\rm{Re}}_D} \nabla^2 \bm{u}.
\end{align}
Here, ${\bm u}=\{u,v\}$ and $p$ are the velocity vector and pressure, respectively, nondimensionalized by the fluid density $\rho^{*}$, the diameter of cylinder $D^{*}$, and the uniform velocity $U_\infty^{*}$.
The asterisk
$(\cdot)^{*}$ 
denotes
dimensional quantities, and the Reynolds number is defined as ${\rm Re}_{D}=U^{*}D^{*}/\nu^{*}$, 
where $\nu^{*}$ is the kinematic viscosity.

\begin{figure}[t]
    \centering
    \includegraphics[width=0.9\linewidth]{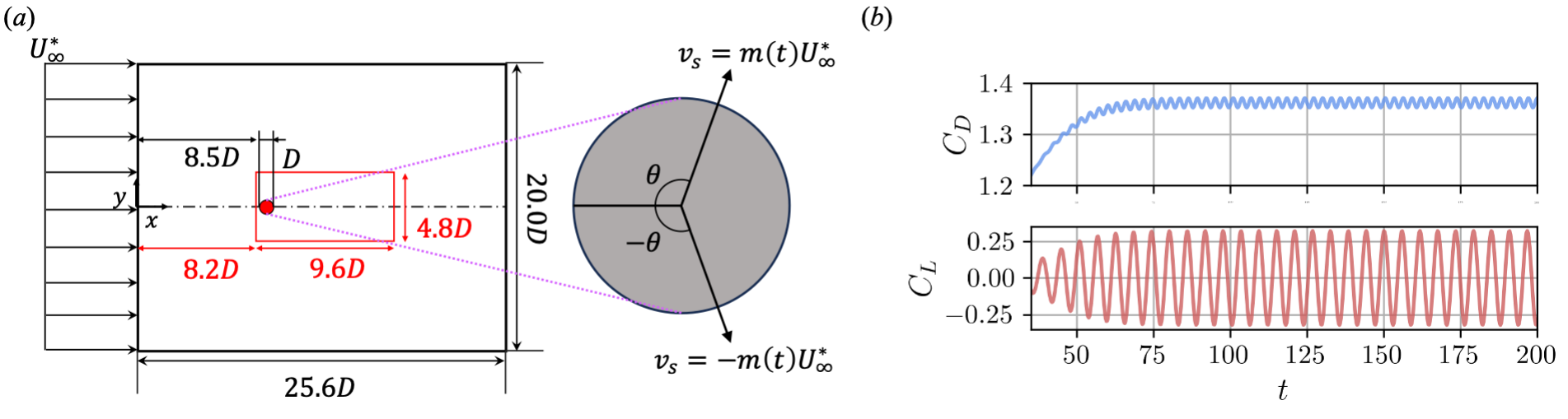}
    \caption{{Problem setting: $(a)$ computational domain of DNS (black), the domain used for machine learning (red), and a schematic of blowing and suction (right); $(b)$ time traces of the drag coefficient ($C_D$) and the lift coefficient ($C_L$) computed by the DNS of the uncont{r}olled case.}}
    \label{fig:domain}
\end{figure}

The left side of Figure~\ref{fig:domain} $(a)$ shows the computational domain, which is defined as a rectangle with the streamwise $(x)$ and transverse $(y)$ dimensions of $25.6D$ and $20.0D$, respectively. 
The center of the cylinder is located at a distance of $9D$ from the inflow boundary.
The present DNS code adopts the second-order central finite difference method on the Cartesian coordinates, and the no-slip condition on the cylinder surface is imposed using the immersed boundary method~\citep{DNS_2D}.
The 
number of computational cells 
in the $x$ and $y$ directions is $\left( N_x, N_y\right) = \left(1024, 800\right)$. 
The time step is set to $\Delta t_{\rm DNS}=2.5\times10^{-3}$.
Free-slip conditions are applied to the upper and lower boundaries, a uniform velocity $U_{\infty} = 1$ is imposed at the inflow boundary, and a convective boundary condition is utilized at the outflow boundary.
For more details of the present DNS and its verification and validation, readers are referred to \citet{DNS_2D}.

As depicted on the right side of Figure~\ref{fig:domain} $(a)$, the locations for blowing and suction are determined following \citet{park1994feedback}, specifically at $\theta = \ang{110}$ from the stagnation point. 
Time-varying blowing and suction is applied in the radial direction with $v_s(t) = m(t)U_{\infty}$, where $m(t)$ represents the magnitude of the blowing and suction.
Blowing is applied to the upper slot and suction to the lower slot when $m(t) > 0$, while the opposite holds when $m(t) < 0$. 
The tangential distribution of the velocity within the blowing/suction slot is assumed Gaussian, such that the amplitude decreases to $1\%$ of its peak at an angle of $\pm \ang{25}$ from the center.

To construct ML models, we utilize the velocity components $\left(u, v\right)$ within the region delineated by the red line in Figure~\ref{fig:domain}, which allows us to concentrate on the flow around the cylinder.
The dimensions of the instantaneous data are $\left(384 \times 192 \times 2\right)$, {\it i.e.}, $384 \times 192$ computational cells in $x$, $y$ directions and two velocity components ($u$, $v$) on each computational point.
As the dataset for the ML model, we generate two different time series datasets of flow fields.
The first dataset captures the periodic wake behind the cylinder with steady blowing and suction.
The magnitude of blowing and suction is expressed as $m(t) = 0, \pm 0.1, \pm 0.3,\pm 0.5, \pm 0.7,\pm 1.0$. 
Note that $m(t) = 0$ is 
the
base flow without blowing and suction.
We utilize data within the time range of {$t = [500, 1000]$ ({\it i.e.}, around 80 cycles of vortex shedding)} with a time interval of $\Delta t_{\rm data}=2.5\times10^{-1}$.
The second dataset focuses on the transient evolution of the cylinder wake.
To emphasize the wake's development, we employ data within the time span of $t = [35, 350]$, with a time interval of $\Delta t_{\rm data}=2.5\times10^{-2}$.
{
Note that the present DNS is started at $t=0$ with the mean flow as the initial condition, and $t=35$ approximately corresponds to the time instant when the first vortex reaches the outflow boundary of the computational domain.
As shown in Figure~\ref{fig:domain} $(b)$, the time histories of the lift and drag coefficients show a continuous transition from a transient process to a steady state.
Each dataset is divided into $70\%$ training data and $30\%$ validation data.
Note that
 there is no need to prepare test data different from the training dataset
 because 
 the purpose of the present work is to extract the linear system
 in the latent space, and 
 what is needed to test the model is to
 perform numerical integration of the extracted linear system from 
 the initial latent variables.
}

\subsection{Partially {non}linear, linear system extraction autoencoder (pn-LEAE)}
\subsubsection{Autoencoder}
\label{subsubsec:AE}
\begin{figure}[b]
    \centering
    \includegraphics[width=0.7\linewidth]{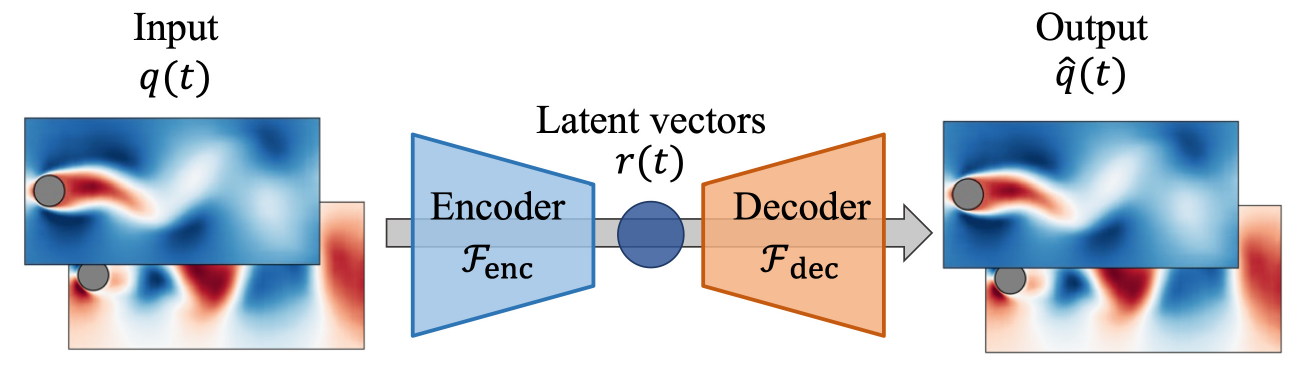}
    \caption{Schematic of the autoencoder.}
    \label{fig:AE}
\end{figure}
Before delving into the methods we propose, it is important to provide an explanation on the standard autoencoder.
Typically, an autoencoder is used to extract essential features from extensive datasets. The diagram of a typical autoencoder is depicted in Figure~\ref{fig:AE}. The autoencoder consists of two components: an encoder $(\cal{F}_{\rm enc})$ and a decoder $(\cal{F}_{\rm dec})$. 
The encoder's function is to transform the input data ${\bm q}$ into a different representation called latent variables or latent vectors, denoted as ${\bm r}$. 
Meanwhile, the decoder's role is to expand this representation to reconstruct the original data, $\hat{\bm q}$, from the latent vector ${\bm r}$. 
These relationships can be expressed as 
\begin{equation}
    \bm{r} = \cal{F}_{\rm enc}(\bm{q})\in \mathbb{R}^{\mathit n_{\mathit r}},\\
    \label{eq:AE-enc}
\end{equation}
\begin{equation}
    \hat{\bm q} = \cal{F}_{\rm dec}(\bm{r})\in {\mathbb R}^{\mathit n_{\mathit q}},
    \label{eq:AE-dec}
\end{equation}
where $n_r$ represents the dimension of the latent vector, and $n_q$ represents the dimension of the original data. 
The encoder and decoder are designed using neural network architectures such as Multi-Layer Perceptron (MLP)~\citep{MLP} and Convolutional Neural Network (CNN)~\citep{LBBH1998}. Their internal weights ${\bm w}$ are fine-tuned during training to ensure that the output data $\hat{\bm q}$ match the input data $\bm{q}$, {\it i.e.},
\begin{align}
    {\bm w}={\rm argmin}_{{\bm w}}{\parallel \bm{q}-\hat{\bm q} \parallel}_2.
\end{align}
If the original data is successfully reconstructed, it implies that the latent vector ${\bm r}$ can be seen as an alternative representation of the original data. 
When the dimensions of the latent vectors are smaller than those of the original data, the autoencoder can be employed for developing a reduced-order model (ROM). 
A typical autoencoder{, whose input and output are identical,} only reduces the size of the data and does not consider the temporal relationship of 
time-series data{, while there are also some attempts to utilize the autoencoder-like network structure
to train the temporal development by considering different time instants for the input and output ({\it e.g.,} \cite{Fukami-MLTG,wang2024})}.

\subsubsection{Linear ordinary differential equation (LODE) layer}
\label{subsubsec:LODE}
Let us explain the details of the linear ordinary differential equation layer (LODE layer) we introduce in the present study. 
{This customized layer is used to numerically integrate a set of linear ordinary differential equations (ODE), and it contains trainable parameters to  identify the system.} 
We are dealing with a system governed by a linear ODE represented as
\begin{align}
    \dot{\bm r}={\bm A}{\bm r} + {\bm B}{\bm m},
    \label{eq:ODE}
\end{align}
where ${\bm r} \in \mathbb{R}^{\mathit n_{\mathit r}}$ is the state of the system and ${\bm m} \in \mathbb{R}^{\mathit n_{\mathit m}}$ is the control input.
{The system matrix ${\bm A}$ and the control matrix ${\bm B}$ are not prescribed but identified by the machine learning.}
For instance, time integration of the left-hand-side of equation~(\ref{eq:ODE}) with the Crank-Nicolson scheme yields
\begin{align}
    {\bm r}(t+\Delta t)&=(2{\bm I}-\Delta t{\bm A})^{-1}
    [(2{\bm I}+\Delta t{\bm A}){\bm r}(t)+\Delta t {\bm B} ({\bm m}(t)+{\bm m}(t+\Delta t)],
    \label{eq:CN_ODE2}
\end{align}
{where ${\bm I}$ denotes the identity matrix.}
\noindent
The LODE layer calculates ${\bm r}(t+\Delta t)$ from ${\bm r}(t)$ and $\Delta t$ by the operations like equation~(\ref{eq:CN_ODE2}).
{The structure of the LODE layer is the same as a two-layer fully-connected linear NN; however, unlike a general NN,}
${\bm A}$ and ${\bm B}$ are set as trainable parameters and 
optimized by training.

{In the present study, we limit our attention to the simplest case, {\it i.e.}, a low Reynolds number, two-dimensional (2D) flow around a circular cylinder
accompanying periodic vortex shedding.
A previous study~\citep{murata2020nonlinear} demonstrated that the 2D flow around a circular cylinder 
at ${\rm Re}_D=100$ can be mapped to a binary latent vector exhibiting periodic oscillations.
Therefore,} assuming that the dynamics of the latent vector is described by a single oscillator, the matrix ${\bm A}$ can be expressed as
\begin{equation}
{\bm A} =
\begin{bmatrix}
0 & -\omega \\
\omega & 0 \\
\end{bmatrix}
.
\label{eq:A}
\end{equation}
{In this specific problem,}
 the angular frequency $\omega$ is 
 {sole}
 trainable parameter {to determine the system matrix ${\bm A}$}.
The training process is divided into two steps.
First, only the matrix ${\bm A}\ (\omega)$ is trained without control inputs.
Then, only the matrix ${\bm B}$ is trained using the flow field with control inputs, while ${\bm A}$ is fixed.
Eventually, the linear ODE represented by ${\bm A}$ and ${\bm B}$ serves as the governing equations for the temporal evolution of ${\bm r}$.

However, linear systems face challenges when applied to nonlinear flow development{, where the orbital radius of the latent trajectory can be varied. }
To address this,
we introduce an enhanced Linear ODE (LODE) layer 
that can handle varying
orbital radius. 
The idea for this enhanced LODE is borrowed from  
the error in 
the growth rate of the radius
when the non-conservative time integration schemes are used. 
For instance, let us consider the oscillator at the unit angular frequency and the state ${\bm r}$ at time $t$ given as
\begin{equation}
{\bm A} =
\begin{bmatrix}
0 & -1 \\
1 & 0 \\
\end{bmatrix}
, \;\;
{\bm r}(0)=
\begin{bmatrix}
\cos{\theta}\\
\sin{\theta}\\
\end{bmatrix}
.
\label{eq:ic_SV}
\end{equation}
The state ${\bm r}$ and the orbital radius $R$ 
at $t+\Delta t$ computed using the Crank-Nicholson method 
are
\begin{align}
{\bm r}(t+ \Delta t) &= \frac{1}{4+\Delta t^2}
\begin{bmatrix}
(4-\Delta t^2)\cos{\theta}-4\Delta t \sin{\theta}\\
4\Delta t \cos{\theta}+(4-\Delta t^2)\sin{\theta}\\
\end{bmatrix}
,\\
    R(t+ \Delta t) &= \parallel{\bm r}(t+ \Delta t)\parallel_{2}=1.
\label{eq:OR_CN}
\end{align}
In this case, the orbital radius $R$ is unchanged, consistent with its mathematical property. 
On the other hand, when the Euler explicit method is used, we can easily compute that the orbital radius changes to $R(t+\Delta t)=\sqrt{1+\Delta t^{2}}$.
It can also easily be shown that 
the desired rate of change in the orbital radius, {\it i.e.}, $\gamma \simeq (R(t+\Delta t)-R(t))/\Delta t$, can be obtained by time integration using the Euler explicit method by replacing the time step $\Delta t$ by a new parameter $\beta$ defined as
\begin{equation}
\beta = -\frac{2\gamma}{R(t)^{2}-\gamma^{2}}.
\label{eq:OR_growth_rate}
\end{equation}
{By}
adding
{such a}
term representing
the growth rate 
to the Crank-Nicholson method, we obtain a
time integration scheme
 that can freely accommodate the evolution of the state, {\it i.e.},
\begin{align}
    {\bm r}(t+\Delta t) &= {\bm r}(t)+\Delta t {\bm A}\left(\frac{{\bm r}(t+\Delta t)+{\bm r}(t)}{2}\right){+}\beta{\bm A}\left(\frac{{\bm r}(t+\Delta t)-{\bm r}(t)}{2}\right),
\end{align}
which can be explicitly expressed as
\begin{align}
        {\bm r}(t+\Delta t)&=(2{\bm I}-(\Delta t +\beta){\bm A})^{-1}(2{\bm I}+(\Delta t-\beta){\bm A}){\bm r}(t).
\end{align}
{Note that the control input has been omitted for brevity.}

In the training process, 
$\gamma$ 
{in equation (\ref{eq:OR_growth_rate})}
is set as a trainable parameter during flow development.
{As a result, $\beta$ in equation (15) is non-zero for transient flows}, while $\beta=0$ in steady periodic state.
More concretely,
the time integration performed by the LODE layer is summarized as
\begin{equation}
\begin{cases}
{\bm r}(t+\Delta t)&=(2{\bm I}-(\Delta t +\beta){\bm A})^{-1}(2{\bm I}+(\Delta t-\beta){\bm A}){\bm r}(t) \hspace{1cm} (t<t_{b}),\\
{\bm r}(t+\Delta t)&=(2{\bm I}-\Delta t{\bm A})^{-1}(2{\bm I}+\Delta t{\bm A}){\bm r}(t)\hspace{2.85cm} (t_{b}\leq t),
\end{cases}
\label{eq:OR_set_beta}
\end{equation}
where $t_{b}$ is the prescribed boundary between the transient and the steady state. 

\subsubsection{Overview of partially nonlinear LEAE}

\begin{figure}[b]
    \centering
    \includegraphics[width=0.7\linewidth]{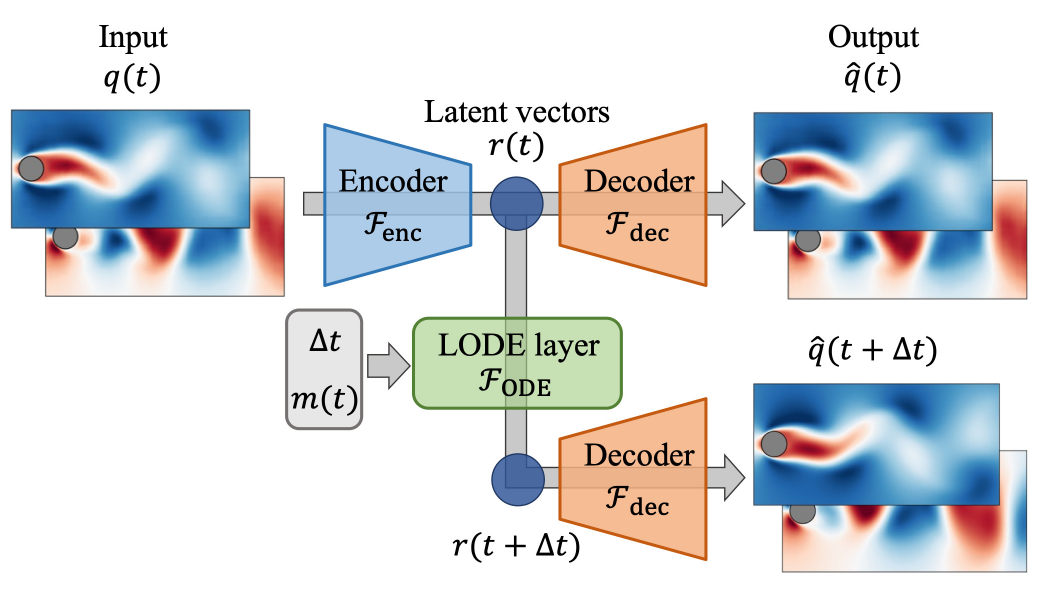}
    \caption{Schematic of the partially nonlinear linear system extraction autoencoder (pn-LEAE).}
    \label{fig:LEAE}
\end{figure}

In this section, we 
overview 
the
partially nonlinear LEAE (pn-LEAE), which is 
built
by combining the methods introduced in Sections~\ref{subsubsec:AE} {and} \ref{subsubsec:LODE} .
The schematic of the pn-LEAE is shown in Figure~\ref{fig:LEAE}.
The model is simultaneously 
trained for 
reconstruction and time evolution, and the weights 
in the two
decoders are shared.
The operations with time evolution by the LODE layer are expressed as
\begin{align}
\begin{cases}
    {\bm r}(t) &= \mathcal{F}_{\rm enc}({\bm q}(t)),\\
    {\bm r}(t+\Delta t) &= \mathcal{F}_{\rm ODE}({\bm r}(t),\Delta t,m(t)),\\
    \hat{\bm q}(t+\Delta t) &= \mathcal{F}_{\rm dec}({\bm r}(t+\Delta t)),\\
    \hat{\bm q}(t) &= \mathcal{F}_{\rm dec}({\bm r}(t)),
\end{cases}
\label{eq:trainP}
\end{align}
where $\Delta t$ is the time interval, $m(t)$ is the control input (a scalar in this case).
The encoder and decoder are trained concurrently to compress low-dimensional features into values that align with a linear system, while also 
{retaining}
reconstruction accuracy.
{
The training process of the pn-LEAE is divided into the following three steps.
For each training step, we use the mean squared error (MSE) as the loss function.
\begin{enumerate}
    \item The AE and the LODE are simultaneously trained using the datasets for both transient and steady periodic flows 
    without control input.
    The key technique here is to train them with the ground truths at two consecutive time instants as shown in Figure \ref{fig:LEAE}, whereby the AE retains its capability as the AE, and the weights in the LODE are trained simultaneously.
    Since the control input is absent at this step, the system equation is expressed as
    $\dot{\bm r}={\bm A}{\bm r}$.
    The matrix ${\bm A}$ and the growth rate ${\bm \gamma}$ are obtained
    as the trainable parameters in addition to the weights in the AE.
    \item An additional training is made using the steady periodic flow data
     by addinig a regularization term to the loss function, {\it i.e.}, 
    \begin{equation}
    L_{\mathrm dis}=\sum(\max{|{\bm r}(t)|}-\min{|{\bm r}(t)|}),
    \label{eq:reg}
\end{equation}
where $L_{\rm dis}$ is distance from origin.
    This fine tuning is introduced to make sure that the latent dynamics adheres to a single oscillation in the case of steady periodic flow, 
    and the implementation of such a constraint allows us to avoid extrapolation problems in the encoder and decoder.
    \item We extend the linear system to $\dot{\bm r}={\bm A}{\bm r}+{\bm B}{\bm m}$ for the application of control inputs, with treating only the matrix ${\bm B}$ as a trainable parameter (namely, matrix ${\bm A}$ and the AE weights are fixed).
\end{enumerate}
}



The network structure of the present pn-LEAE is basically similar to the CNN-AE used by \citet{murata2020nonlinear}, with the differences that the aforementioned LODE layer is inserted in the bottleneck part and that the latent vector is decoded by two decoders {(sharing the weights)} for two consecutive time steps as illustrated in Figure~\ref{fig:LEAE}.
{There are also some minor differences in the structures in the encoder and decoder parts, {\it i.e.}, the use of a multi-scale model and batch normalization layers.
Interested readers are referred to Appendix for the detailed structures.}

\subsection{Control design for linear system}
\label{subsec:method_LQR}
We employ the Linear Quadratic Regulator (LQR) \citep{LQR} 
{and the ${\cal H}_\infty$ control approach \citep{Doyle1989}}
to formulate a state feedback control law for the extracted linear systems. 
LQR is a methodology used to devise an optimal control strategy for systems governed by linear ODEs while minimizing costs. 
{In addition, ${\cal H}_\infty$ control considers external disturbances and
is designed to minimize the ${\cal H_\infty}$ norm of the transfer function \citep{Zhou1988}.
Among different solution procedures, here we follow the Riccati equation-based procedure \citep{Bewley2001} which can describe the LQR and the ${\cal H}_\infty$ approach in a unified manner.}

{Our state-feedback system (\ref{eq:ODE})
under external disturbances ${\bm w}$ can be represented as
\begin{align}
    \dot{\bm r}=&{\bm A}{\bm r} + {\bm B}{\bm m} + {\bm B_1}{\bm w},\\
    \bm z=&{\bm C}{\bm r} + {\bm D}{\bm m},
    \label{eq:ODEw}
\end{align}
where $\bm z$ is the controlled output \citep{Zhou1988}.
By setting ${\bm C}={\bm I}$ and  ${\bm D}=\ell \bm {I}$, the quadratic cost function $J$ for the latent vector ${\bm r}$ and the control input ${\bm m}$ 
is represented as
\begin{equation}
    J = \int_{0}^{\infty} \left[ {\bm r}^T{\bm r}+\ell^2{\bm m}^{T}{\bm m}
    -\Gamma^2 {\bm w}^{T}{\bm w}
    \right]dt,
\end{equation}
where $\ell$ and $\Gamma$ represent the costs of control and disturbance, respectively \citep{Bewley2001}.
In this formulation, the LQR is regarded as the special case of ${\cal H}_\infty$ controllers where $\Gamma \rightarrow \infty$.}

The control input ${\bm m}$ to minimize the $J$ is derived as
\begin{equation}
{\bm m}(t) = {\bm {K}}\bm {r}(t),
\label{eq:min_J}
\end{equation}
{
where $\bm{K}$ is the feedback gain, i.e.,
\begin{equation}
{\bm K} = -{\frac{1}{\ell^2}}{\bm {BP}},
\label{eq:gain}
\end{equation}
and}
${\bm P}$ is the target matrix
obtained
by solving the {corresponding} Riccati 
equation,
\begin{equation}
{
    {\bm A}^{T}{\bm P}+\bm{PA}-
    \bm{P}\left(
    \frac{1}{\ell^2}\bm{B}{\bm B}^{T}-\frac{1}{\Gamma^2}\bm{B}_1\bm{B}_1^{T}\right){\bm P} +{\bm I} = 0.
    \label{eq:Ricatti}
    }
\end{equation}

Note that, in the present study, both the system matrix ${\bm A}$
and the control matrix ${\bm B}$ are identified by machine learning using the LEAE.
{
For the disturbance matrix, a na\"ive choice, $\bm{B}_1 = {\bm I}$, is made. 
}

\section{Results and discussion}
\subsection{Linear system extraction}
\label{subsec:LE}
\begin{figure}[b]
    \centering
    \includegraphics[width=0.95\linewidth]{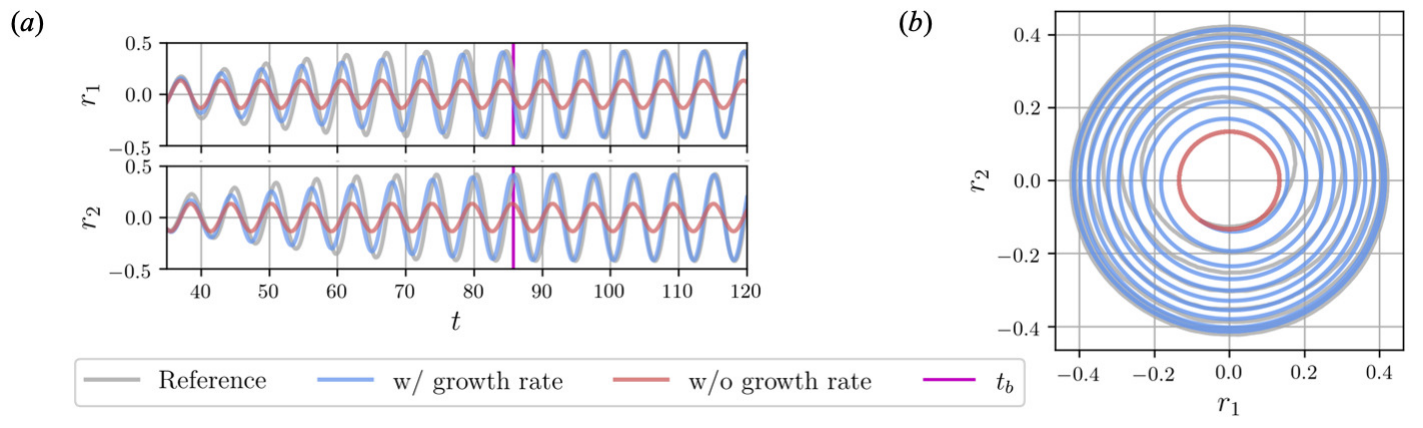}
    \caption{{Dynamics of the latent vector: 
    $(a)$ time trace of each latent variable, drawn with   
    the boundary between transient and steady periodic states, $t_b$; $(b)$ trajectory of the latent vector. ``Reference'' represents the dynamics of latent vector simply extracted by the CNN-AE, and ``w/ growth rate'' and ``w/o growth rate'' represent those computed by time integration of the system identified by the pn-LEAE, with and without introducing the growth rate.}}
    \label{fig:LV_base}
\end{figure}
\begin{figure}[t!]
    \centering
    \includegraphics[width=1\linewidth]{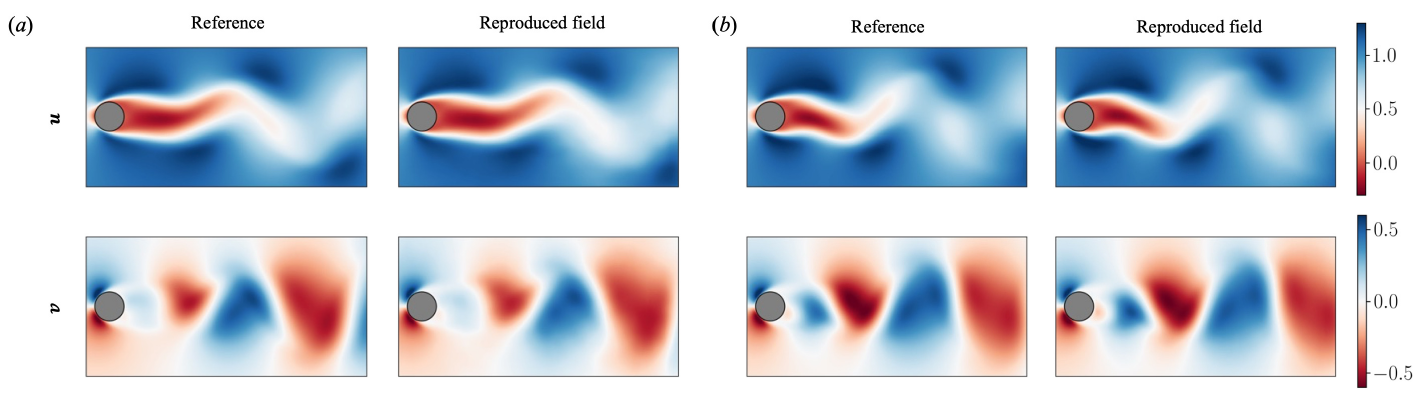}
    \caption{Comparison between the reference DNS and the flow fields reproduced by the pn-LEAE: $(a)$ $t=40$ (transient process); $(b)$ $t=100$ (steady periodic state).}
    \label{fig:res_rep}
\end{figure}
\begin{figure}[b!]
    \centering
    \includegraphics[width=0.8\linewidth]{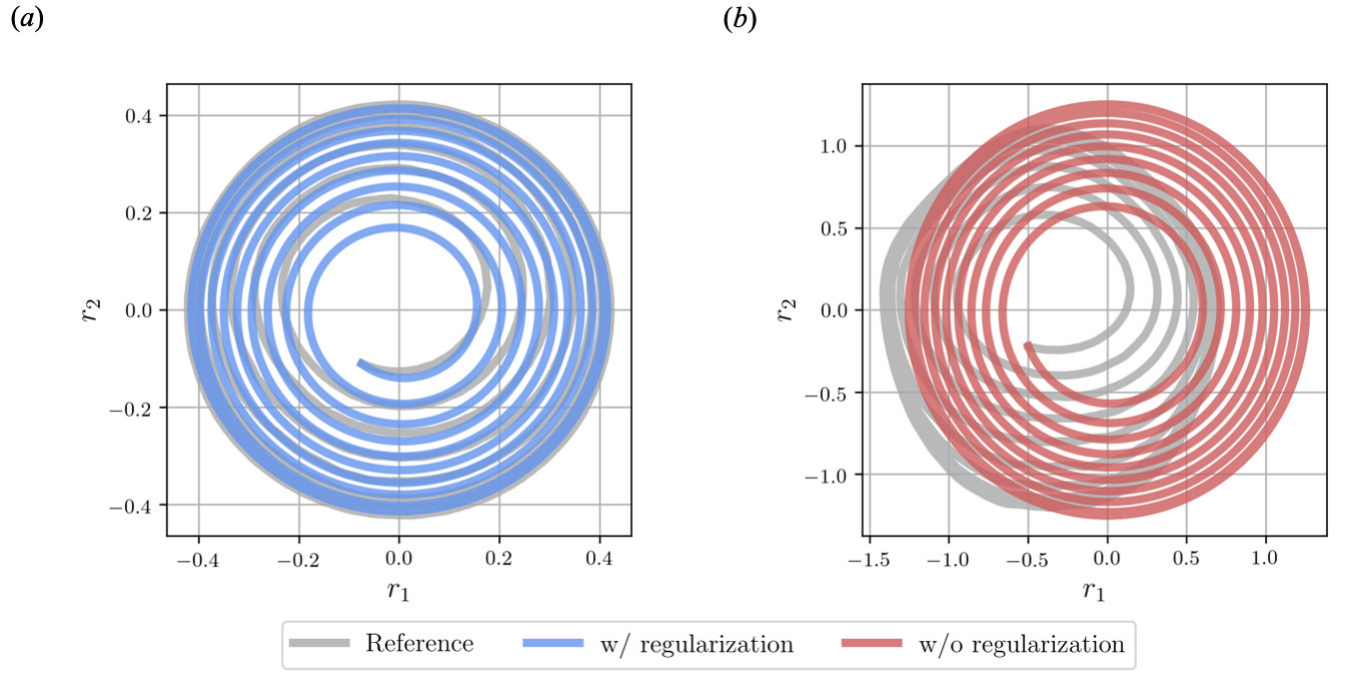}
    \caption{{Comparison of 
     the trajectories of the latent vectors computed from the linear systems identified with and without regularization term: $(a)$ with regularization term; $(b)$ without regularization term.}}
    \label{fig:res_constraint}
\end{figure}
\begin{figure}[b!]
    \centering
    \includegraphics[width=0.95\linewidth]{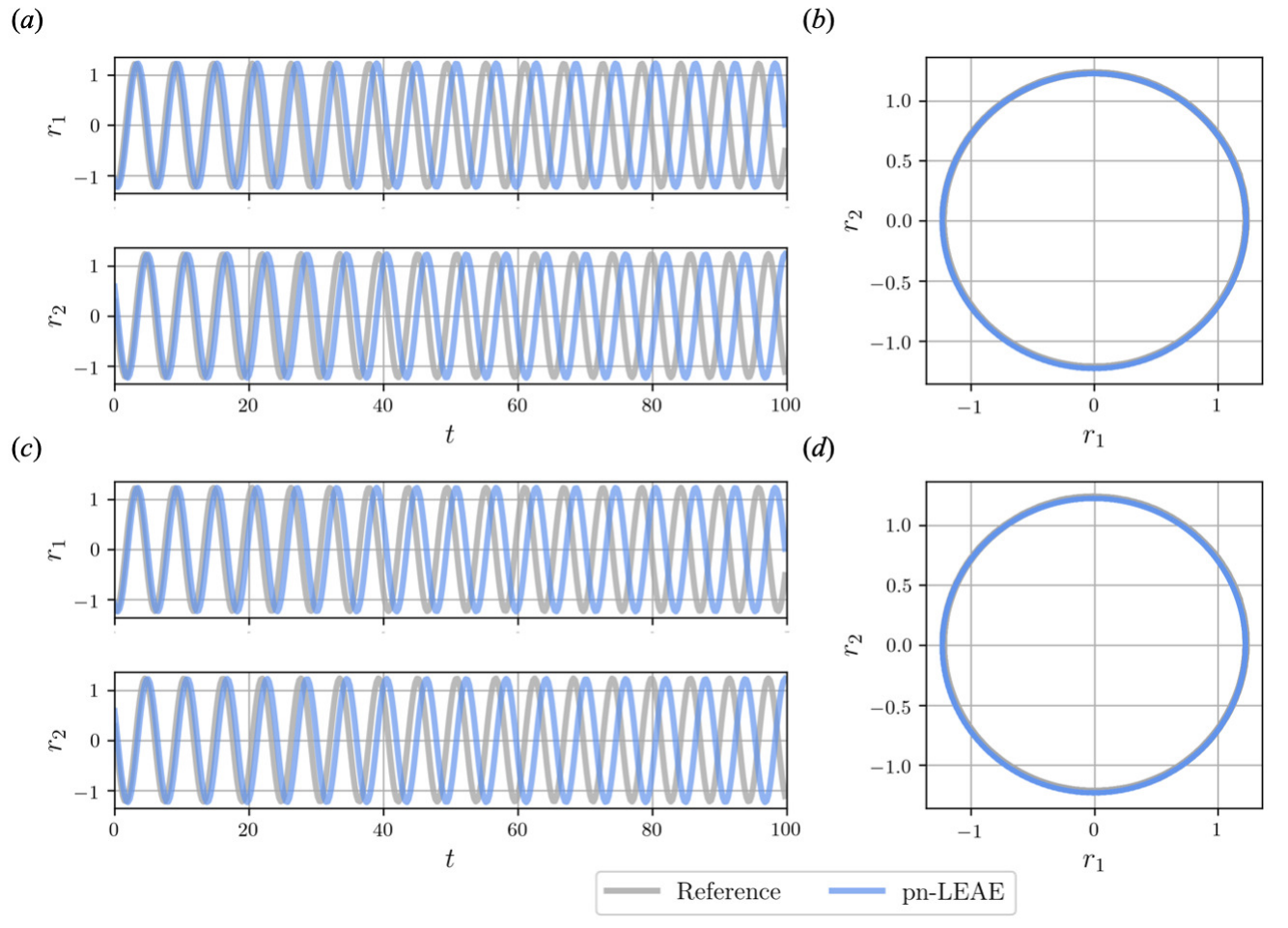}
    \caption{Dynamics of the latent vector in the presence of steady blowing and suction: 
    $(a)$ time trace of each latent variable, $m(t)=0.3$; $(b)$ trajectory of the latent vector, $m(t)=0.3$;
    $(c)$ time trace of each latent variable, $m(t)=1.0$; $(d)$ trajectory of the latent vector, $m(t)=1.0$.}
    \label{fig:res_traj_cont}
\end{figure}
\begin{figure}[t!]
    \centering
    \includegraphics[width=1\linewidth]{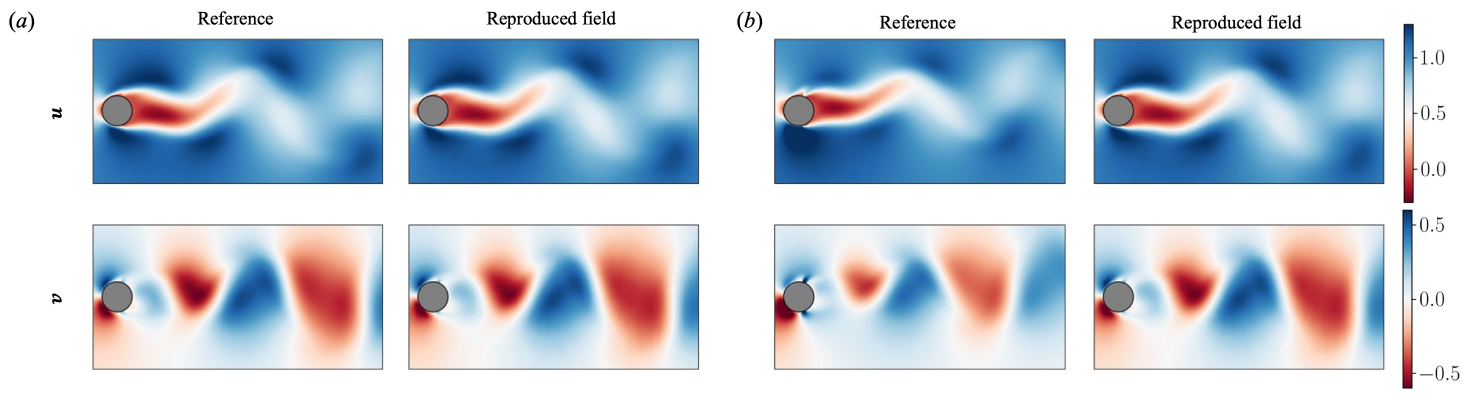}
    \caption{Comparison between the reference DNS and the flow fields reproduced by the pn-LEAE in the presence of steady blowing and suction: $(a)$ $m(t)=0.3$; $(b)$ $m(t)=1.0$.}
    \label{fig:res_rep_cont}
\end{figure}

In this section, we focus on the linear system extraction 
by the proposed
pn-LEAE and discuss its performance.
First, we demonstrate its applicability to a flow field involving a transition process.
The linear 
ODE
governing
 the latent dynamics obtained by the present pn-LEAE is
\begin{align}
\dot{\bm r}(t) =
\begin{bmatrix}
0 &-1.066\\
1.066& 0\\
\end{bmatrix}
{\bm r}(t),
\label{eq:Linear_system}
\end{align}
with the growth rate 
in the transient period ($t<t_b$) of
 $\gamma=1.332\times10^{-3}$.

Figure~\ref{fig:LV_base} shows the predicted results of the temporal evolution of the latent vectors using the extracted ODEs, along with the boundary distinguishing between the time intervals where the growth rate is considered or not.
{In this figure, ``Reference'' represents the dynamics of latent vector simply extracted by the CNN-AE, and ``w/ growth rate'' and ``w/o growth rate'' represent those computed by a time integration of equation~ (\ref{eq:Linear_system}) with the initial condition ({\it i.e.}, ``Reference'' at the initial time).}
The prediction results align reasonably well with the reference, and the model can accommodate both the evolution of the latent vectors and the steady periodic oscillation.
In particular, focusing on the period from the transient state to the steady periodic state, we observe that the evolution of the latent space comes to a halt, and a gradual transition takes place towards the steady periodic behavior. 
{Figure~\ref{fig:LV_base} also demonstrates the difference in  predictions with and without introducing the growth rate. 
Without introducing the growth rate, the trajectory of the latent vector keeps a constant radius and does not express the transient behavior.}
These results also suggest that the behavior of the latent vectors can be described by a simple linear ODE as expressed in equation (\ref{eq:Linear_system}) with the help of the additional trainable parameter $\gamma$ introduced. 

The flow field 
reconstructed by the decoder from the predicted latent vectors
is
depicted in Figure~\ref{fig:res_rep}. 
Here, we compare two states:
 $(a)$ the
transient state 
 and $(b)$ the steady state. 
Both flow fields closely align with the DNS data, highlighting the suitability of the temporal evolution of the latent vector as described by the extracted linear system. 
This finding is a crucial indicator of the model's effectiveness.

{
Additionally, we discuss the advantages of the regularization term proposed in Equation~(\ref{eq:reg}). 
Figure~\ref{fig:res_constraint} shows the trajectories of the latent vectors
computed from the linear systems identified with and without the regularization term. 
Without regularization,
the trajectory of
the latent vector 
is observed to 
traverse to the right.
This traversal 
will
cause
an extrapolation 
issue.
On the other hand, the trajectory with the regularization is observed to be within the space covered by ``Reference'' ({\it i.e.,} the training data).
Therefore, we recognize that this regularization term plays a crucial role in preventing extrapolation issues and improving the mapping performance to a linear system.
}

Then, we investigate the applicability of the 
{model trained}
with prescribed constant control inputs, {\it i.e.}, {$m(t)= \{0, \pm 0.1, \pm 0.3, \pm 0.5, \pm 0.7, \pm 1.0\}$, assuming the situation that the time scale of the feedback control is sufficiently longer than the time step used for the time integration of the linear system, as will be explained in Section 3.3}.
Here, we extract a linear system using equation~(\ref{eq:CN_ODE2}) 
while fixing the
weights of matrix ${\bm A}$.
In other words, 
the training 
optimizes 
only matrix ${\bm B}$ with respect to the control inputs.
The obtained linear system with 
the constant
control inputs is
\begin{align}
\dot{\bm r}(t) =
\begin{bmatrix}
0 &-1.066\\
1.066& 0\\
\end{bmatrix}
{\bm r}(t)+
\begin{bmatrix}
{-0.03672}\\
{0.02672}\\
\end{bmatrix}
{m}(t).
\label{eq:Linear_system_wcont}
\end{align}
Figure~\ref{fig:res_traj_cont} shows the prediction results of the latent vectors based on the flow field under two different control input magnitudes: $m(t)=0.3$ and $1.0$. 
{Again, ``Reference'' represents the dynamics of latent vector simply extracted by the CNN-AE, and ``pn-LEAE'' represents those obtained by integrating equation (\ref{eq:Linear_system_wcont}).}
While 
a slight phase shift 
is
observed for both control inputs, the temporal change in the orbital radius 
is well captured, showcasing the effectiveness of the present linear system. 
The phase-shift problem is an inherent challenge arising from the accumulation of small errors within the linear system and remains an important subject for future research.
{
However, we believe that this phase-shift issue should have little impact on the effectiveness of the control law as far as the local relationship in the latent space is reasonably well represented.
}

In Figure~\ref{fig:res_rep_cont}, we observe the reconstructed fields alongside the reference data obtained from DNS. 
Both predictions demonstrate the response of to the steady blowing and suction,
 indicating that the model effectively 
 capture the
 flow development, including control effects, through a straightforward addition within the latent space.

\subsection{Design of control laws using the derived linear system}

In this section, we elucidate the method used to derive control laws from the linear system obtained in Section~\ref{subsec:LE}.
We also employ optimal control, following the process described in Section~\ref{subsec:method_LQR}.
Specifically, we design a state feedback control law for the linear system described by equation~(\ref{eq:Linear_system_wcont}) using LQR
{and ${\cal H_\infty}$ control approaches}. 
It is essential to note that the rank of the controllability matrix
is equal to $2$, which corresponds to $n_r$ in this system, indicating its controllability.
{The cost parameters $\ell^2$ and $\Gamma^2$ are chosen as shown in Table \ref{eG}}
such that the resultant values of control input $m(t)$ do not deviate significantly from the values used for training.
Solving equation~(\ref{eq:Ricatti}) 
by 
using
the
Schur decomposition method,
we can obtain
{
the feedback gain $\bm{K}$ as shown in Table~\ref{eG}.
Note that even
larger $\ell^2$ and smaller $\Gamma^2$ were not considered because they result in even smaller feedback gain.
Also, under the present values of $\ell^2$, the feedback gain $\bm{K}$ in ${\cal H_\infty}$ control converges to those in the LQR control already at $\Gamma^2 = 10000$.
}

\begin{table}[t!]
{
\begin{center}
    \caption{Parameters $\ell^2$ and $\Gamma^2$ in the LQR and ${\cal H}_\infty$ control, the obtained feedback gain, $\bm{K}=-(1/\ell^2)\bm{BP}$, and the resultant RMS lift fluctuations ($C^\prime_{L}$) normalized by the uncontrolled value ($C^\prime_{L0}$) when applied to the DNS introduced in Section 3.3.}
    \begin{tabular}{cccccc}
    \hline
     \hline
        Controller & $\ell^2$ & $\Gamma^2$ & $\bm{K}$ & $C^\prime_{L}/C^\prime_{L0}$ \\
     \hline
        LQR & 10 & $\infty$ & $(0.3629, -0.2614)$ & 0.43 \\
        LQR & 3 & $\infty$ & $(0.6655, -0.4755)$ & 5.06 \\
        ${\cal H}_\infty$ & 10 & 1000 & $(0.1224, -0.08942)$ & 1.03 \\
        ${\cal H}_\infty$ & 3 & 1000 & $(0.4757, -0.3508)$ & 0.43 \\
	\hline
     \hline
    \end{tabular}
    \label{eG}
\end{center}
}
\end{table}

\begin{figure}[t!]
    \centering
    \includegraphics[width=1\linewidth]{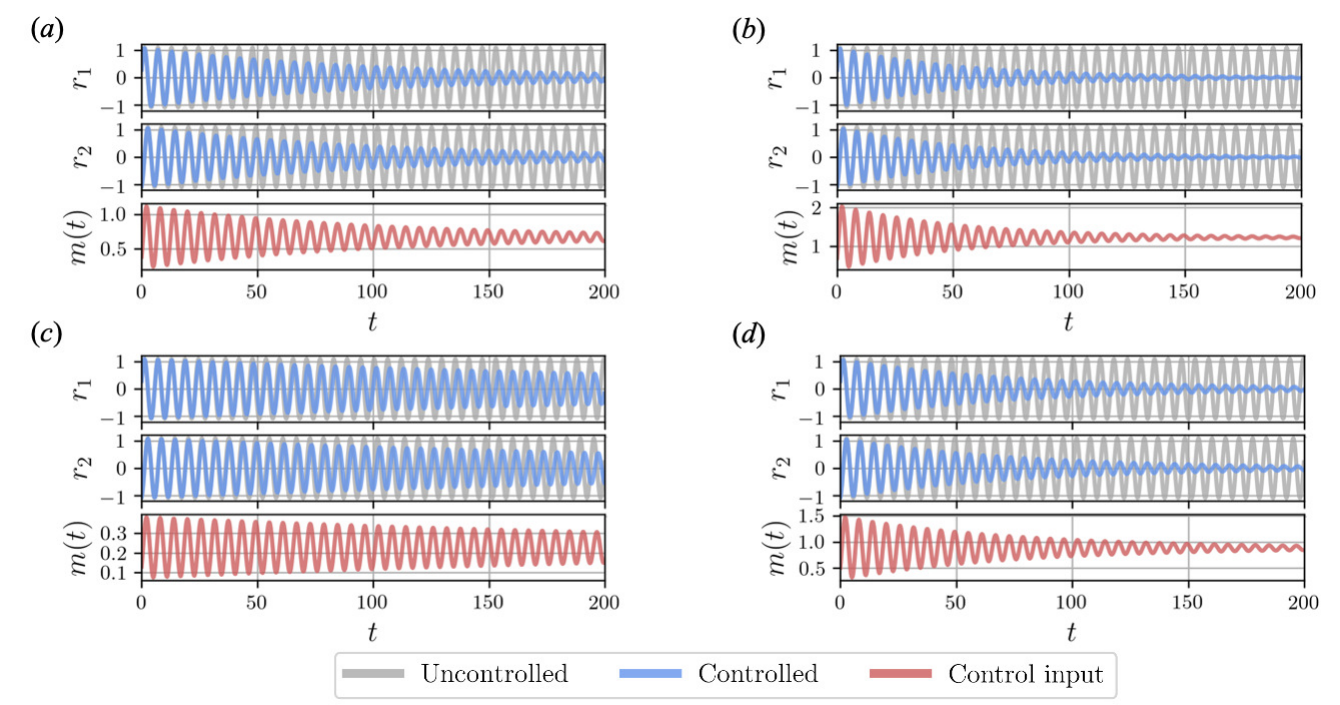}
    \caption{Control effect when the 
    {controller}
    is applied in the latent space: 
    {
    $(a)$ LQR with $\ell^2 = 10$; 
    $(b)$ LQR with $\ell^2 = 3$; 
    $(c)$ ${\cal H}_\infty$ with $(\ell^2, \Gamma^2)=(10,1000)$;
    $(d)$ ${\cal H}_\infty$ with $(\ell^2, \Gamma^2)=(3,1000)$.
    }
    }
    \label{fig:res_latent_cont}
\end{figure}
\begin{figure}[t!]
    \centering
    \includegraphics[width=1\linewidth]{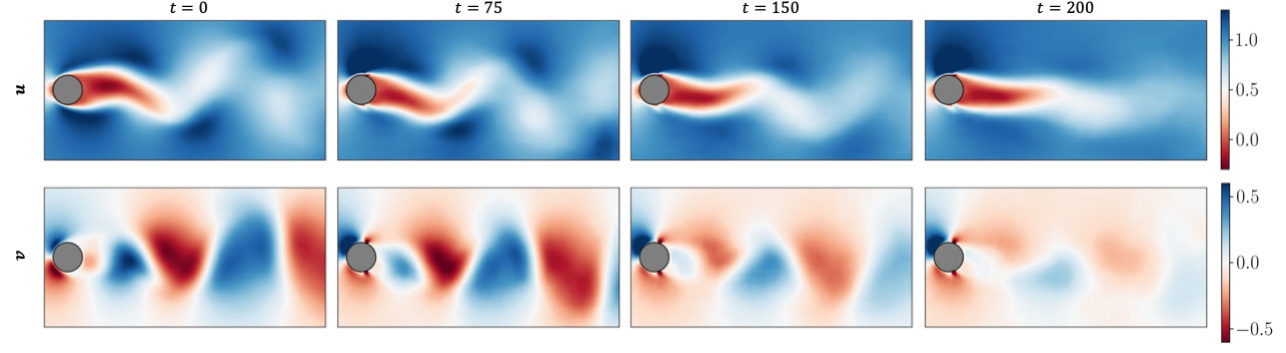}
    \caption{Flow fields reconstructed from the latent dynamics controlled by the LQR regulator 
    {with $\ell^2=10$}, {\it i.e.}, equation~(\ref{eq:Linear_system_wcont}).}
    \label{fig:res_rep_cont_bylatent}
\end{figure}
\begin{figure}[t!]
    \centering
    \includegraphics[width=0.8\linewidth]{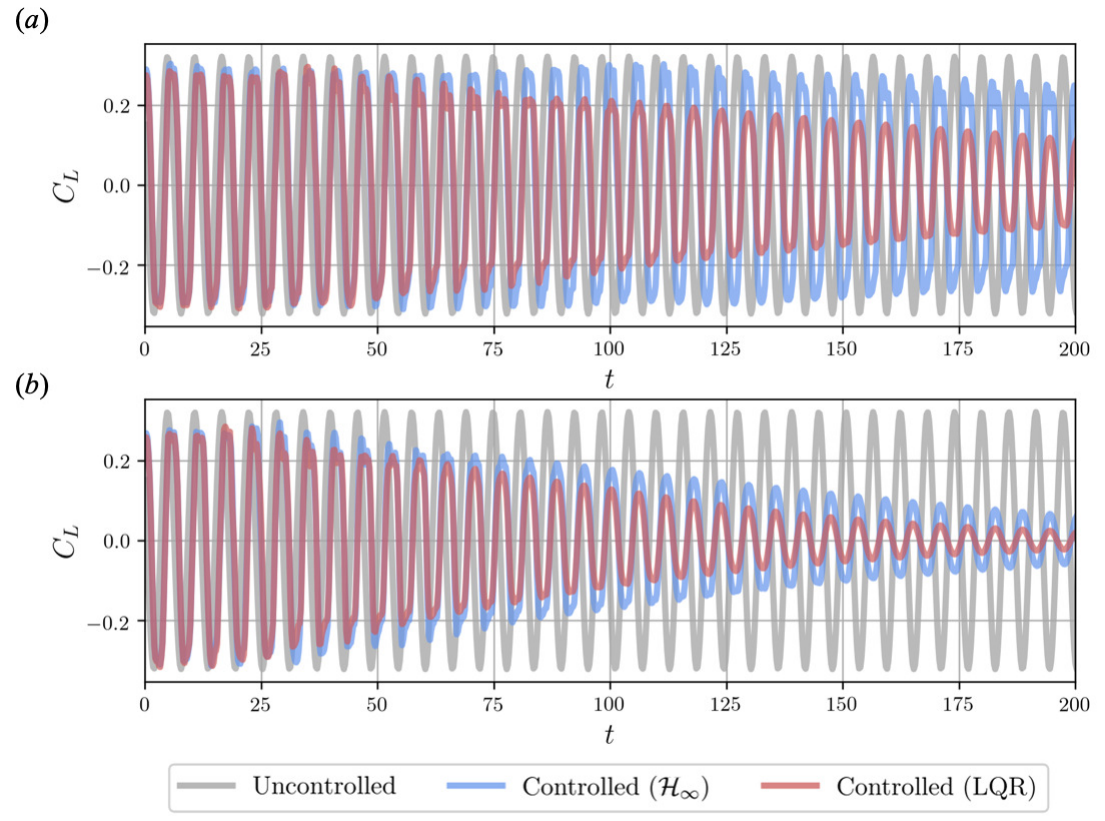}
    \caption{{Control effect on the lift coefficient when the 
    {controller}
    is applied in the latent space. 
    {
    ($a$) LQR and ${\cal H}_{\infty}$ with $\ell^2=10$;
    ($b$) LQR and ${\cal H}_{\infty}$ with stronger control input, {\it i.e.}, $\ell^2=3$.}
    The values in the uncontrolled case are directly computed from the DNS data using the control volume method, while those in the controlled cases are computed from the decoded fields using the control volume method.}}
    \label{fig:CL_CV_LQRcont}
\end{figure}

Before implementing this control law
 in DNS, we first apply this feedback control law to the linear system
  to confirm its performance within the latent space.
The control 
results
for ${\bm r}$ 
{by the LQR controller with $\ell^2=10$}
are displayed in Figure~\ref{fig:res_latent_cont}($a$). 
The state converges to the origin as a result of the control intervention. 
{T}he control input $m(t)$ is found to fall within the range of $[-1.0, 1.0]$, which is within the maximum and minimum values of the control input seen in the training data.
For comparison, the result obtained with 
{the other controllers}
are also presented in  Figure~\ref{fig:res_latent_cont}($b$)--($d$).
{The amplitudes of the latent variables and the control input decay faster for the cases with a larger feedback gain. 
However, the initial control amplitude in the case of LQR with $\ell^2=3$ ({\it i.e.}, $m\simeq 2$) significantly deviates the range used for training ({\it i.e.}, $|m|\leq 1$).
This implies that this case may not work for the actual flow, and it should be verified in the next section by coupling with DNS.}

Figure~\ref{fig:res_rep_cont_bylatent} shows the flow fields reconstructed from the time histories of {\bm r}. 
It is evident that this behavior in the latent space corresponds to the suppression of vortex shedding in 
the physical space. 
As a result, we can also expect that this control law will successfully suppress vortex shedding when applied to DNS.

{
To quantitatively
demonstrate the 
control performance,
the time history of the lift coefficient
is presented in Figure~\ref{fig:CL_CV_LQRcont}.
As clearly observed, the lift fluctuations are significantly suppressed by the present control --- the suppression rate amounts over 90\%
for 
{the case of LQR with $\ell^2=3$}.
}

\subsection{Application of designed feedback control law to DNS}
\begin{figure}[t]
    \centering
    \includegraphics[width=0.9\linewidth]{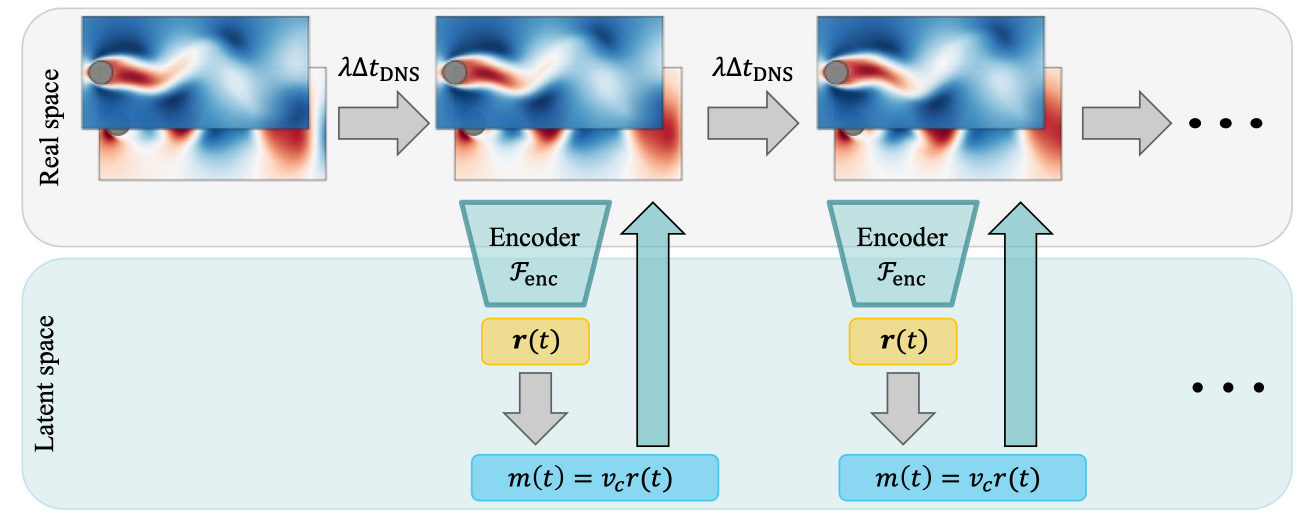}
    \caption{Schematic diagram of feedback control using the extracted control law.}
    \label{fig:overview_DNS_FBcont}
\end{figure}
\begin{figure}[t]
    \centering
    \includegraphics[width=1\linewidth]{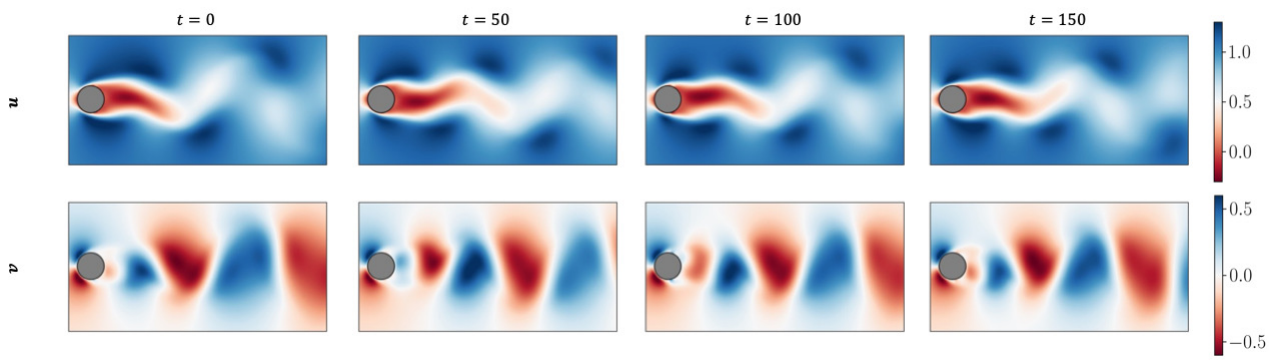}
    \caption{Flow fields computed by DNS with the present feedback control law {(LQR with $\ell^2=10$)}.    
    }
    \label{fig:res_DNS}
\end{figure}
\begin{figure}[t]
    \centering
    \includegraphics[width=1\linewidth]{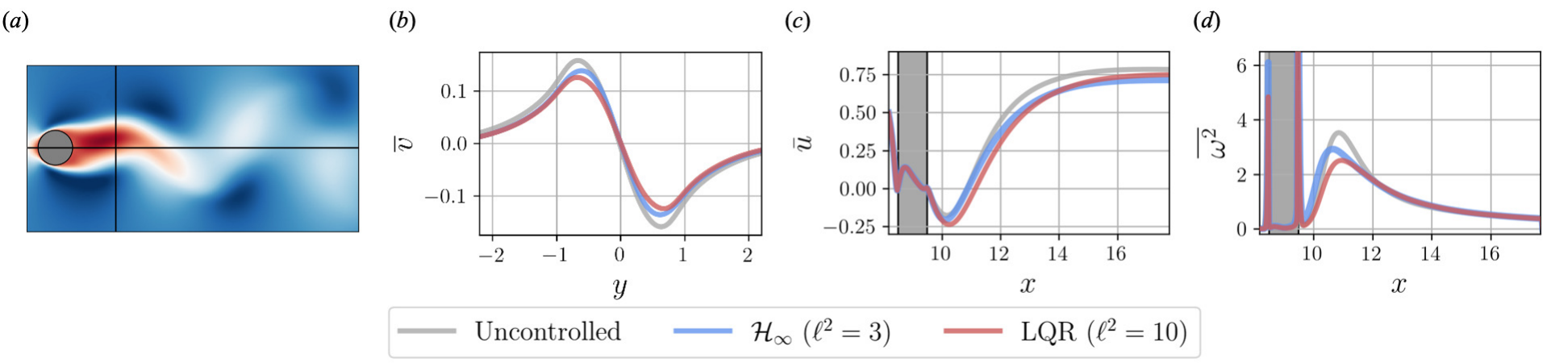}
    \caption{{
    Fundamental flow statistics: 
    ($a$) illustration of measurement planes (vertical line: $1.5D$ downstream of the cylinder; horizontal line, centerline);
    ($b$) mean transverse velocity at $1.5D$ downstream of the cylinder;    
    ($c$) mean centerline velocity, 
    ($d$) vorticity fluctuations along the centerline.}}   
    \label{fig:res_stat}
\end{figure}
\begin{figure}[t]
    \centering
    \includegraphics[width=1\linewidth]{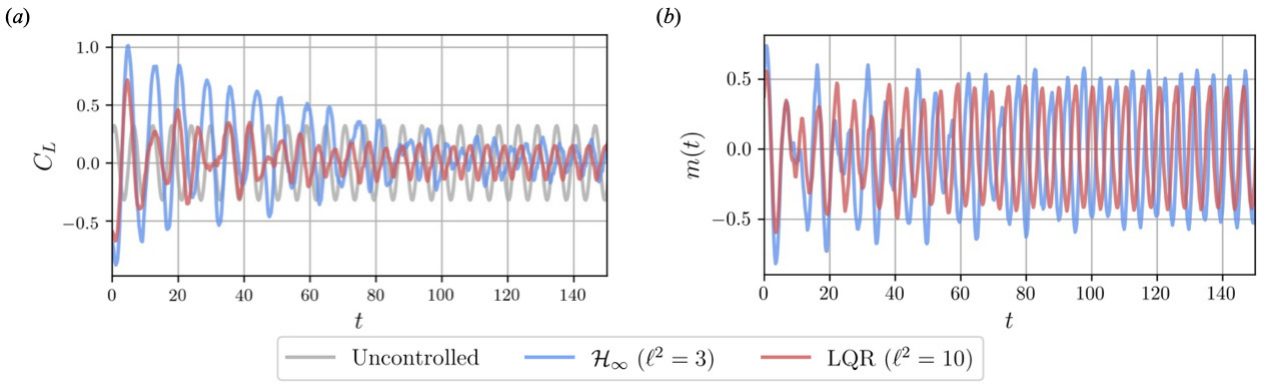}
    \caption{Time traces computed by DNS: $(a)$ lift coefficient $C_L$ without control and with designed feedback control; $(b)$ control input $m(t)$.}
    \label{fig:res_CL_mt}
\end{figure}
Using  
the derived control law 
({\ref{eq:min_J}}), the control input in the DNS
is 
expressed as
\begin{align}
m(t) = {\bm{K}}
{\cal F}_{\rm enc}({\bm u}(t)),
\label{eq:cont_law_DNS}
\end{align}
where $m(t)$ represents the intensity of blowing and suction in the physical space,
and ${\bm u}(t)$ is the velocity field computed by DNS.
We apply this feedback control law to the DNS, keeping the DNS configuration consistent with Section~\ref{subsec:DNS}.

The procedure 
of the feedback control applied to the DNS 
is presented in Figure~\ref{fig:overview_DNS_FBcont}.
DNS is performed 
in the physical space
with
the control input 
updated at 
prescribed
time intervals ($\lambda$ time steps). 
{
The time interval for applying feedback control was set to $\lambda=100$ time steps, which corresponds to approximately 1/25 of the vortex shedding period. 
}
In the latent space, the flow field is mapped to latent vectors using the trained encoder, and the control input is determined by applying the control law (\ref{eq:cont_law_DNS}).
Repeating these procedures is expected to suppress vortex shedding. 
A crucial aspect of this method is the connection between the latent space and the real space through a control law represented 
by
binary values.

{As shown in Table~\ref{eG}, the parameters for the control laws are the same as those in Section 3.2.
Table~\ref{eG} also presents
the resultant root-mean-square (RMS) lift fluctuations ($C^\prime_{L}$) normalized by the uncontrolled value ($C^\prime_{L0}$).
}
{
While the lift fluctuations are reduced over 50\% in the cases of LQR with $\ell^2=10$ and 
${\cal H}_{\infty}$ with $(\ell^2,\Gamma^2)=(3,1000)$, the most successful case in the latent space ({\it i.e.}, LQR with $\ell^2=3$) increases the lift fluctuations.
This is likely due to the large initial control amplitude outside the training range, as has been concerned in Section 3.2.
For ${\cal H}_{\infty}$ with $(\ell^2,\Gamma^2)=(10,1000)$, the lift fluctuations are nearly unchanged, which
may be attributed to the insufficient control amplitude.
Therefore, in the following, we focus on the two successful cases, i.e., LQR with $\ell^2=10$ and ${\cal H}_{\infty}$ with $(\ell^2,\Gamma^2)=(3,1000)$.
}

Figure~\ref{fig:res_DNS} shows the DNS 
results 
with the control input
{obtained by the LQR with $\ell^2=10$.}
Over time, there is a noticeable reduction in vortex shedding, indicating the success of the feedback control with the designed control law.
{
Figure~\ref{fig:res_stat} shows the control effect to the 
fundamental flow statistics, {\it i.e.,} 
the mean transverse velocity at $1.5D$ downstream of the cylinder (Figure~\ref{fig:res_stat}($b$)),  
the mean centerline velocity (Figure~\ref{fig:res_stat}($c$)), 
and the vorticity fluctuations along the centerline (Figure~\ref{fig:res_stat}($d$)){, in the cases of LQR with $\ell^2=10$ and ${\cal H}_{\infty}$ with $(\ell^2,\Gamma^2)=(3,1000)$}.
As clearly observed, all of these statistics
show 
modifications consistent with the suppression of
vortex shedding. 
}

The control effect is also evident from the lift coefficient depicted in Figure~\ref{fig:res_CL_mt} $(a)${, which indicates about 50\% sup{p}ression in lift fluctuations.}
Nevertheless, the issue of imperfect control effectiveness persists.
{Although the ${\cal H}_{\infty}$ control looks to cope with the phase shift better than the LQR, the resultant reduction in the lift fluctuations is similar to that of the LQR as presented in Table~\ref{eG}.} 
We are confronted with the challenge that the control effect plateaus and does not further improve. 
The ultimate objective of this study is to completely eliminate vortex shedding{, as has been observed in the latent space.}
Analyzing the physical quantities and control inputs presented in Figure~\ref{fig:res_CL_mt}, it becomes apparent that after $t=80$, both remain within a specific range. 
To address this challenge, it would be necessary to enhance the diversity of training data when training autoencoders. 
In the future, research efforts should focus on devising strategies to improve control effectiveness while operating under various constraints.

\section{Conclusions}
In this study, we 
{proposed
a partially nonlinear linear-system extraction autoencoder (pn-LEAE), which consists of convolutional neural networks-based autoencoder (CNN-AE) and a custom layer to extract a low-dimensional latent dynamics
from fluid velocity field data.
This pn-LEAE is designed to extract a linear dynamical system so that the modern control theory can easily be applied, while a nonlinear compression is done with the autoencoder (AE) part so that the latent dynamics conforms to that linear system.
The key technique is to train this pn-LEAE with the ground truths at two consecutive time instants, whereby the AE part retains its capability as the AE, and the weights in the linear dynamical system are trained simultaneously.}
The proposed LEAE has shown that a low-Reynolds number flow around a circular cylinder at ${\rm Re}_D=100$
can be effectively described by a simple equation representing a single oscillation.
Furthermore, we presented a partially nonlinear LEAE (pn-LEAE), an enhanced version of LEAE capable of handling transient and feedback-controlled flows.
A pn-LEAE has demonstrated its ability to generalize and adapt to flow evolution by incorporating a growth rate of the orbital radius.
Finally, we designed a control law based on the linear systems extracted by LEAE and provided evidence of its effectiveness. 
This outcome underscores the capacity of the designed control law to bridge the gap between latent space and real space.

{
Unlike the other successful ML approaches for flow control such as the reinforcement learning \citep{rabault2019artificial,Castellanos2022,Pino2023}, the merit of the present approach is that we can {more} directly utilize the control theory.
Although the flow problem dealt with in the present study is limited to an extremely fundamental one, {\it i.e.,} a low-Reynolds number two-dimensional flow around a circular cylinder with periodic vortex shedding, 
and the control th{eo}ry applied is also limited to the standard LQR {and an ${\cal H}_{\infty}$ control,}
we believe that the basic concept of the present work can be extended to deal with more complex flow problems
by relaxing the constraint on the system matrix, increasing the size of the system matrix so that multiple oscillating frequencies can be considered, and so on.
For instance, it has been demonstrated that a low-Reynolds number turbulent channel flow can be well represented by $\sim 1000$ POD or AE modes \citep{Nakamura2021,Racca2023}. 
Also, a substantial part of the mechanism can be represented by oscillating wave motions at different frequencies, as extensively studied in the resolvent analysis \citep{McKeon2010}. 
The remaining challenge will be to approximate the remaining nonlinear interactions (corresponding to the forcing term in the resolvent analysis) using the AE part.
From these, we believe that it should be possible, in principle, to apply the present approach to a low-Reynolds number fully developed turbulence by adopting $\sim 1000$ latent variables, although additional difficulties accompanied by the increase in the matrix size can also be expected.
The ultimate goal is to develop a machine-learning-based system linearizer akin to the Koopman operator.
Such extensions remain open for the future research.
}

\section*{Appendix}
\begin{figure}[t]
    \centering
    \includegraphics[width=0.5\linewidth]{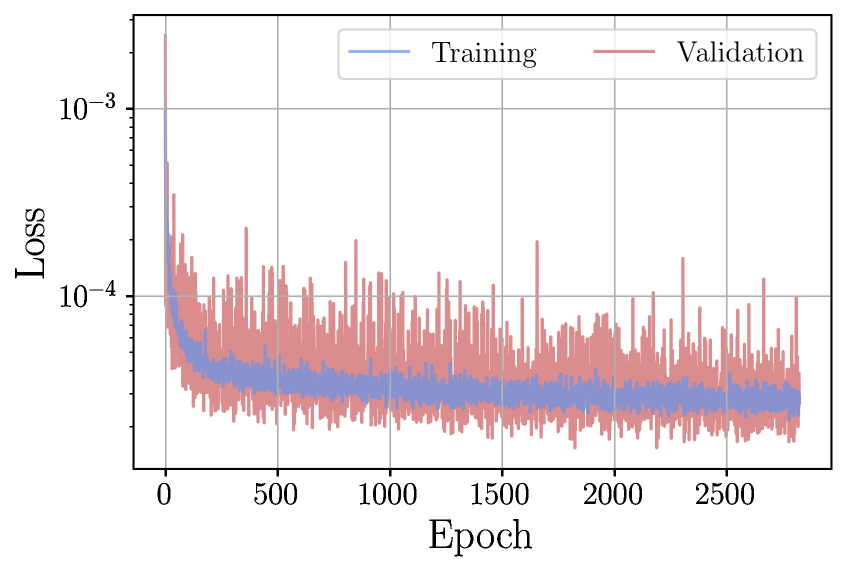}
    \caption{{The typical error curve in the training process of pn-LEAE.}}
    \label{fig:app_lossc}
\end{figure}

{In this Appendix, we provide some additional information on the present machine learning. 
Table~\ref{tab:arch_enc} and Table~\ref{tab:arch_dec} present the structures of the encoder and the decoder, respectively. 
Figure~\ref{fig:app_lossc} exemplifies the evolution of the loss function during the typical training process of pn-LEAE.}
\begin{table}[t]
    \centering
    \caption{{Architecture of the encoder composing the LEAE. A multi-scale model with multiple filter sizes is employed from 1st Convolution to 7th Maxpooling.}}
    \begin{tabular}{cccc}
    \hline
     \hline
        Layer & Output shape &{F}ilter sizes & Activation \\
        \hline
         \hline
        Input & $(384\times192\times2)$& &\\
                \hline 
        1st Convolution& $(384\times192\times16)$&((3,3),(5,5),(9,9)) &tanh\\
        Batch normalization& $(384\times192\times16)$& \\
        2nd Convolution& $(192\times96\times8)$&((3,3),(5,5),(9,9)) &tanh \\
        Batch normalization& $(192\times96\times8)$& \\
        2nd Maxpooling& $(96\times48\times8)$& \\
        3rd Convolution& $(96\times48\times8)$&((3,3),(5,5),(9,9)) &tanh \\
        Batch normalization& $(96\times48\times8)$& \\
        3rd Maxpooling& $(48\times24\times8)$& \\
        4th Convolution& $(48\times24\times8)$&((3,3),(5,5),(9,9)) &tanh \\
        Batch normalization& $(48\times24\times8)$& \\
        4th Maxpooling& $(24\times12\times8)$& \\
        5th Convolution& $(24\times12\times8)$&((3,3),(5,5),(9,9)) &tanh \\
        Batch normalization& $(24\times12\times8)$& \\
        5th Maxpooling& $(12\times6\times8)$& \\
        6th Convolution& $(12\times6\times4)$&((3,3),(5,5),(9,9)) &tanh \\
        Batch normalization& $(12\times6\times4)$& \\
        6th Maxpooling& $(6\times3\times4)$& \\
        7th Convolution& $(6\times3\times4)$&((3,3),(5,5),(9,9)) &tanh \\
        Batch normalization& $(6\times3\times4)$& \\
        7th Maxpooling& $(2\times1\times4)$& \\
        \hline
        Add& $(2\times1\times4)$& \\
        8th Convolution& $(2\times1\times4)$&(5,5) &tanh \\
        Batch normalization& $(2\times1\times4)$& \\
        Reshape& (8)& \\
        Dense & (2)& & Linear\\
        \hline
         \hline
    \end{tabular}
    \label{tab:arch_enc}
\end{table}
\begin{table}[t]
    \centering
    \caption{{Architecture of the decoder composing the LEAE. A multi-scale model with multiple filter sizes is employed from 1st Upsampling and 8th Convolution and Batch normalization.}}
    \begin{tabular}{cccc}
    \hline
     \hline
        Layer & Output shape & {F}ilter size{s} & Activation \\
        \hline
         \hline
        Input & $(2)$& &\\
        Dense& $(8)$&&tanh\\
        Reshape& $(2\times1\times4)$& \\
        1st Convolution& $(2\times1\times4)$&(5,5) &tanh \\
        Batch normalization& $(2\times1\times4)$& \\
        \hline 
        1st Upsampling& $(6\times3\times4)$& \\
        2nd Convolution& $(6\times3\times4)$&((3,3),(5,5),(9,9)) &tanh \\
        Batch normalization& $(6\times3\times4)$& \\
        2nd Upsampling& $(12\times6\times4)$& \\
        3rd Convolution& $(12\times6\times4)$&((3,3),(5,5),(9,9)) &tanh \\
        Batch normalization& $(12\times6\times4)$& \\
        3rd Upsampling& $(24\times12\times4)$& \\
        4th Convolution& $(24\times12\times8)$&((3,3),(5,5),(9,9)) &tanh \\
        Batch normalization& $(24\times12\times8)$& \\
        4th Upsampling& $(48\times24\times8)$& \\
        5th Convolution& $(48\times24\times8)$&((3,3),(5,5),(9,9)) &tanh \\
        Batch normalization& $(48\times24\times8)$& \\
        5th Upsampling& $(96\times48\times8)$& \\
        6th Convolution& $(96\times48\times8)$&((3,3),(5,5),(9,9)) &tanh \\
        Batch normalization& $(96\times48\times8)$& \\
        6th Upsampling& $(192\times96\times8)$& \\
        7th Convolution& $(192\times96\times16)$&((3,3),(5,5),(9,9)) &tanh \\
        Batch normalization& $(192\times96\times16)$& \\
        7th Upsampling& $(384\times192\times16)$& \\
        8th Convolution& $(384\times192\times2)$&((3,3),(5,5),(9,9)) &tanh \\
        Batch normalization& $(384\times192\times2)$& \\
        \hline 
        Add&$(384\times192\times2)$\\
        9th Convolution& $(384\times192\times2)$&(5,5) &tanh \\
        \hline
         \hline
    \end{tabular}
    \label{tab:arch_dec}
\end{table}

\section*{Acknowledgments}

This work 
was supported by JSPS KAKENHI Grant Number 21H05007
The authors acknowledge Mr. Shoei Kanehira {and Mr. Riku Goto} (Keio University) 
for fruitful discussion and comments.


\end{document}